\documentclass[%
reprint,
 amsmath,amssymb,
 aps,
]{revtex4-2}
\usepackage{graphicx}% Include figure files
\usepackage{dcolumn}% Align table columns on decimal point
\usepackage{bm}% bold math
\usepackage{float}
\usepackage{caption}
\usepackage{subcaption}
\usepackage{color}
\usepackage{tikz}
\usetikzlibrary{shapes.geometric, arrows}

\tikzstyle{startstop} = [rectangle, rounded corners, minimum width=2cm, minimum height=1cm, text centered, draw=black, fill=red!30]
\tikzstyle{process} = [rectangle, minimum width=2cm, minimum height=1cm, text centered, draw=black, fill=blue!30]
\tikzstyle{decision} = [diamond, minimum width=3cm, minimum height=1cm, text centered, draw=black, fill=green!30]
\tikzstyle{arrow} = [thick,->,>=stealth]

\begin{document}

\preprint{APS/123-QED}
\title{Rogue Wave Statistics from a Sparse Coherent Structure Decomposition}

\author{Yuchen He$^{1}$} 
\author{Amin Chabchoub$^{2,3,4}$}
\email{amin.chabchoub@oist.jp}
\author{Zhan Wang$^{1}$}
\email{zwang@imech.ac.cn}

\affiliation{$^1$ Key Laboratory for Mechanics in Fluid Solid Coupling Systems, Institute of Mechanics, Chinese Academy of Sciences, Beijing 100190, China}
\affiliation{$^2$ Marine Physics and Engineering Unit, Okinawa Institute of Science and Technology, Onna-son, Okinawa 904-0495, Japan} 
\affiliation{$^{3}$ Department of Civil and Environmental Engineering, Imperial College London, London SW7 2AZ, United Kingdom}
\affiliation{$^{4}$ Department of Infrastructure Engineering, The University of Melbourne, Victoria 3010,
Australia}

\date{\today}% It is always \today, today,
          
\begin{abstract}
While conventional rogue wave statistical models rely on linear or weakly nonlinear descriptions of random seas, we demonstrate experimentally that moderately or strongly nonlinear wave fields can be represented by sparse ensembles of coherent soliton-like packets. These packets exhibit log-normal amplitude distributions together with uniformly distributed phases and peak emergence times. This sparse coherent structure framework naturally leads to an extreme value description in which the probability of exceedance is governed by the tail of the coherent-structure amplitude distribution. The prediction is validated against laboratory hydrodynamic experiments across a variety of unidirectional sea states, showing good agreement with the experimental observations and comparing favourably with conventional statistical prediction models while retaining analytical simplicity. Our results provide a direct physics-based link between sparse coherent structures and rogue wave probabilities, with broader implications for nonlinear wave physics in optics, cold gases, and plasmas.
\end{abstract}

\maketitle

%\tableofcontents

Rogue waves (RWs) are spontaneous, exceptionally large ocean waves with elevations exceeding twice the significant wave height. They pose a serious threat to marine operations and coastal infrastructure~\cite{kharif2008rogue,dysthe2008oceanic,dudley2019rogue,mori2023science}. Beyond oceanography, analogous extreme wave phenomena have been identified and extensively studied in other nonlinear dispersive systems, including nonlinear optics and Bose--Einstein condensates~\cite{dudley2019rogue,romero2024experimental}. Despite decades of research, the prediction of RWs remains challenging. Beyond the constraints of operational forecasting, accurately describing the statistics of strongly nonlinear random wave fields remains an important open problem. Conventional statistical models typically represent such fields as superpositions of many linear or weakly nonlinear wave components~\cite{longuet1974breaking,onorato2001freak}. Frameworks based on the classical Rayleigh distribution are restricted to Gaussian like statistics with a fixed kurtosis of three~\cite{mori2006kurtosis}. More advanced weakly nonlinear models incorporate corrections associated with skewness and kurtosis~\cite{longuet1963effect,tayfun1980narrow,tayfun2007wave}, but may still underestimate extreme event probabilities in strongly nonlinear sea states~\cite{onorato2013rogue,toffoli2011extreme,slunyaev2013super}.

A substantial body of theoretical and experimental work has established that four-wave interactions and Benjamin-Feir instability, also known as modulational instability (MI), can lead to heavy-tailed wave statistics~\cite{osborne2010nonlinear,maestrini2026revealing}. Within Zakharov's Hamiltonian framework for water waves, the kurtosis was shown to scale with the square of the Benjamin--Feir index (BFI), thereby enhancing the probability of extreme waves~\cite{janssen2003nonlinear}. Large-scale wave-flume experiments subsequently confirmed that the Rayleigh distribution can underestimate RW occurrence by more than an order of magnitude at large BFI~\cite{onorato2006extreme}. A quantitative relation between kurtosis and freak wave occurrence further connected nonlinear wave statistics to practical risk assessment~\cite{mori2006kurtosis}. 

In parallel, nonlinear spectral diagnostics based on the inverse scattering transform (IST) have provided a complementary, structure-based description of extreme wave dynamics~\cite{shabat1972exact,osborne2010nonlinear,onorato2021observation}. Within the unidirectional nonlinear Schrödinger equation (NLS), RWs have been shown to be deterministically predictable from the characteristic width scales of higher-order envelope soliton-like packets \cite{cousins2016reduced}. Such coherent structures can also provide a basis for data-driven approaches to predicting extreme events \cite{farazmand2019extreme}. Indeed, breather waves \cite{akhmediev1997solitons}, which describe the nonlinear stage of MI, can be regarded as multi-solitons on a finite background \cite{tikan2017universality,chabchoub2021peregrine}. 
Fully nonlinear simulations have demonstrated substantially enhanced kurtosis and RW probability relative to third-order models, with the maximum excess kurtosis scaling approximately linearly, rather than quadratically, with the BFI~\cite{zhai2025fully}. Field observations further indicate that extreme wave group energy and duration follow generalized extreme value distributions, with New Wave theory underestimating extreme group energy by 20--50\% ~\cite{fu2024statistics}. 

In this Letter, we introduce a statistical framework based on sparse and fundamental coherent structures that provides a structural description of RW emergence, grounded in controlled laboratory experiments. We show that the complex wave envelopes of a deep-water random wave field can be decomposed into a sparse ensemble of coherent, localized, solitonic wave packets whose amplitudes follow a log-normal distribution, while their phases and peak emergence times follow uniform distributions. Rather than describing the field as a dense superposition of arbitrary wave components, this representation reveals a statistical ensemble of relatively isolated structures. The resulting exceedance probability follows directly from the tail of their amplitude distribution, yielding a closed form prediction of RW statistics that requires only the soliton amplitude distribution as input. Beyond hydrodynamics, the framework may be applicable to other nonlinear dispersive media in which analogous extreme waves and coherent structures emerge~\cite{dudley2019rogue,chabchoub2026extreme}.

Hydrodynamic experiments were performed in a $25~\text{m}$-long wave flume with a constant water depth of $0.75~\text{m}$. To ensure deep-water conditions, long-crested random waves were synthesized by a piston-type wavemaker following a JONSWAP spectrum with peak frequency $f_p=1.3~\text{Hz}$ and peak enhancement factor $\gamma=3.3$ or $6$, with characteristic wave steepness $\varepsilon=0.08$ and $0.10$ and significant wave height $H_s=0.035$ and $0.044~\text{m}$ \cite{hasselmann1973measurements}. The peak wavenumber is then $k_p=\frac{(2\pi f_p)^2}{g}$, where $g=9.81~\text{m/s}^2$ is the gravitational acceleration. Given the standard deviation $\sigma$ of the surface elevation, the significant wave height is $H_s=4\sigma$, the characteristic wave amplitude is $a_{\operatorname{char}}=\sqrt{2}\sigma$, and the corresponding characteristic steepness is $\varepsilon=a_{\operatorname{char}}k_p$. Resistive wave gauges, sampling at $32~\text{Hz}$, recorded the surface elevation along the flume, see~\cite{he2022evidence} for further details of the experimental setup. The complex envelope $\psi(t)$ of the measured surface elevation $\eta(t)$ is extracted via bandpass filtering around the peak frequency $f_p$, followed by the Hilbert-Huang transform to suppress oscillations in the wave envelope~\cite{osborne2010nonlinear}. To decompose the wave field into constituent coherent structures, we implement an iterative matching pursuit algorithm. At the core of this soliton decomposition is the representation of the envelope as a superposition of localized soliton-like packets
\begin{equation}
\psi(t) = \sum_{j} N_j a_j \operatorname{sech}\left(\frac{\sqrt{2}}{2} a_j k_p \omega_p (t - t_{0,j})\right) e^{i\phi_j},
\end{equation}
where $a_j$, $t_{0,j}$, and $\phi_j$ denote the amplitude, temporal center, and phase of the $j$-th soliton, respectively. The product of the peak wavenumber and peak wave frequency $k_p\omega_p$ scales the width according to the deep-water dispersion relation $\omega_p=\sqrt{gk_p}=2\pi f_p$, ensuring dimensional consistency and the appropriate scaling of the soliton width~\cite{chabchoub2016hydrodynamic,osborne2010nonlinear}, while the continuous shape factor $N_j$ relaxes the integer-order constraint of NLS soliton theory. The corresponding temporal surface elevation is then reconstructed from the real part of the envelope modulated by the carrier wave
\begin{equation}
\eta(t) =\operatorname{Re}\left(\psi(t) e^{-i\omega_p t}\right).
\end{equation}

\begin{figure}[hbp]
\includegraphics[width=0.48\textwidth]{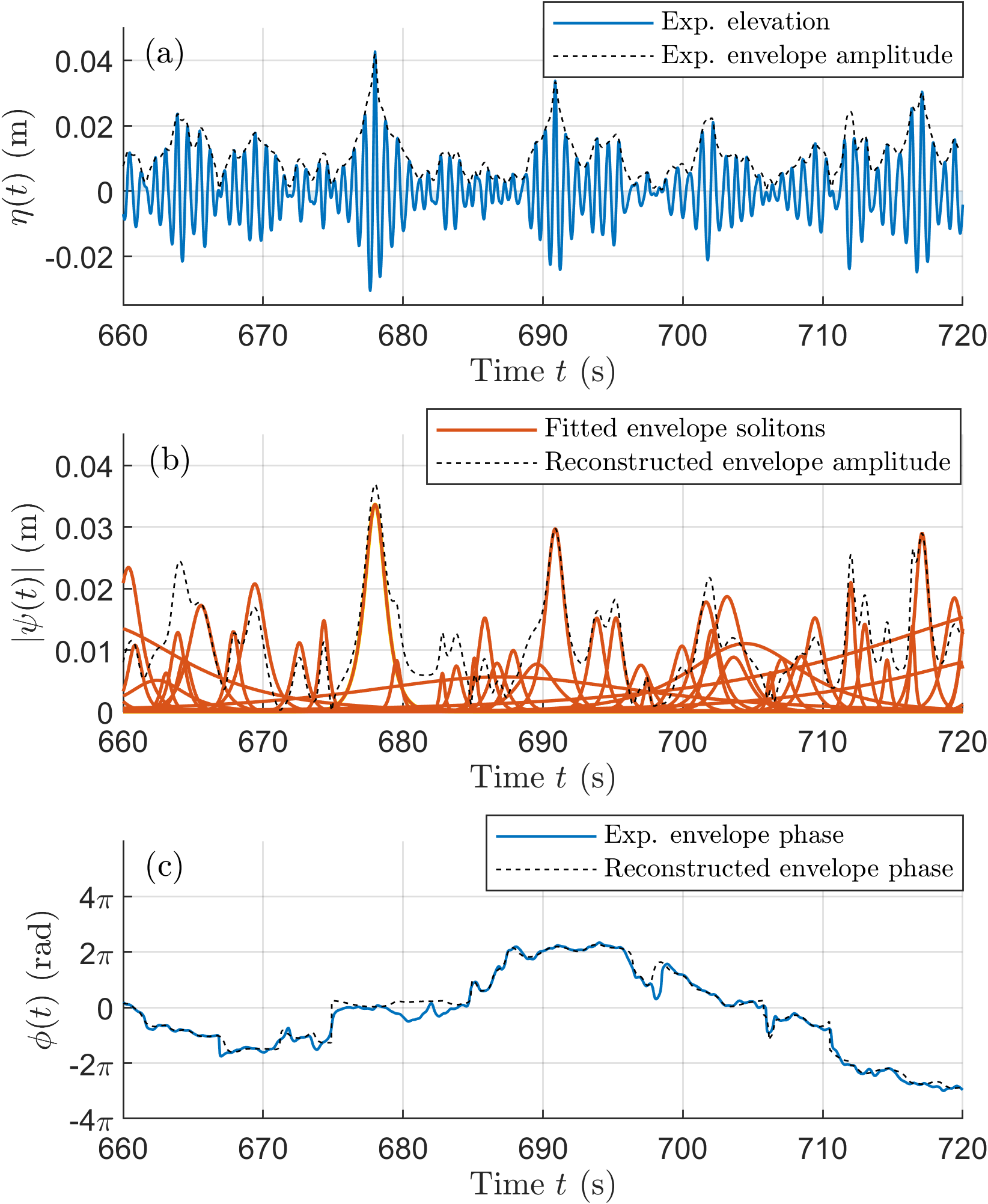}
\caption{Soliton decomposition of a random wave field. (a) Measured surface elevation $\eta(t)$ (blue solid line) and corresponding envelope $|\psi(t)|$ (black dashed line) extracted via the Hilbert-Huang transform. (b) Soliton decomposition of the wave field into a sparse ensemble of individual soliton envelopes (red solid lines) using the matching pursuit algorithm. The total reconstructed envelope (black dashed line) closely follows the experimental profile. (c) Comparison between the experimental data phase (blue solid line) and the soliton-reconstructed phase (black dashed line), demonstrating excellent phase coherence despite the random nature of the field.}
\label{fig1}
\end{figure}

It is instructive to draw an analogy between this soliton decomposition and the more familiar wavelet transform. In a wavelet transform, an arbitrary signal is represented as a linear superposition of translated and dilated copies of a mother wavelet $\psi_w$
\begin{equation}
\eta(t) = \sum_{j} c_j , \psi_w\left(\frac{t - b_j}{a_j}\right),
\end{equation}
where $c_j$ are coefficients, $b_j$ are translation parameters, and $a_j$ are scale parameters. The basis functions are predetermined by the choice of the mother wavelet, limiting the ability of the wavelet decomposition to adapt to the underlying coherent wave structures, as it does not account for complex phase information or nonlinear wave dynamics. Our soliton decomposition follows a similar philosophy but employs a fundamentally different basis. Unlike wavelet bases, the soliton templates are not predetermined; instead, they are extracted iteratively from the signal using a matching pursuit algorithm, see Supplemental Material for details. At each step, the algorithm identifies and extracts the most energetic localized wave packet by fitting the soliton template and minimizing a weighted error over a local time window, using an adaptive grid search over $a$, $N$, and $\phi$, followed by local refinement. After subtracting the dominant soliton from the residual signal, the procedure is repeated until the residual energy drops below $5\%$. The resulting basis functions are therefore adapted to the nonlinear coherent structures present in the wave field rather than imposed by a predetermined mathematical basis. This physics-based representation identifies coherent structures directly, without requiring the IST, and provides the set ${a_j,N_j,\phi_j,t_{0,j}}$ of all identified solitons for the subsequent statistical analysis. As shown below, the ability of this representation to capture the statistical properties of extreme waves suggests that the soliton basis provides a particularly effective description of the underlying nonlinear wave dynamics.

Figure~\ref{fig1} illustrates the soliton decomposition of a representative JONSWAP wave envelope field measurement, where a series of localized wave packets are computed and isolated using the matching
pursuit algorithm. The JONSWAP sea state is parametrized by a peak frequency $f_p=1.3$ Hz, significant height $H_s=0.035$ m, peakedness parameter $\gamma=6$, thus, the characteristic steepness is $\varepsilon=0.08$. Although such a nonlinear soliton decomposition is inherently non-unique, our iterative algorithm yields a consistent and physically robust representation. Remarkably, the superposition of the extracted solitons accurately reproduces the original envelope with a residual error below $5\%$, see an example in Fig.~\ref{fig1}(b), while faithfully capturing the phase evolution, see Fig.~\ref{fig1}(c). These results demonstrate that the seemingly random wave field can fundamentally be interpreted as a coherent superposition of soliton-like structures that preserves both amplitude and phase characteristics. Notably, for this 1000-peak-period record, the soliton decomposition yields only 862 solitons, confirming the sparsity of the representation and supporting the extreme-value framework. Indeed, the typical time interval between adjacent solitons significantly exceeds their individual widths. This separation implies that the solitons evolve primarily via localized, pairwise interactions with neighboring coherent structures rather than through collective coupling with the entire background wave field, a feature that underpins the subsequent statistical analysis.

\begin{figure*}[htp]
\centering
\includegraphics[width=0.975\textwidth]{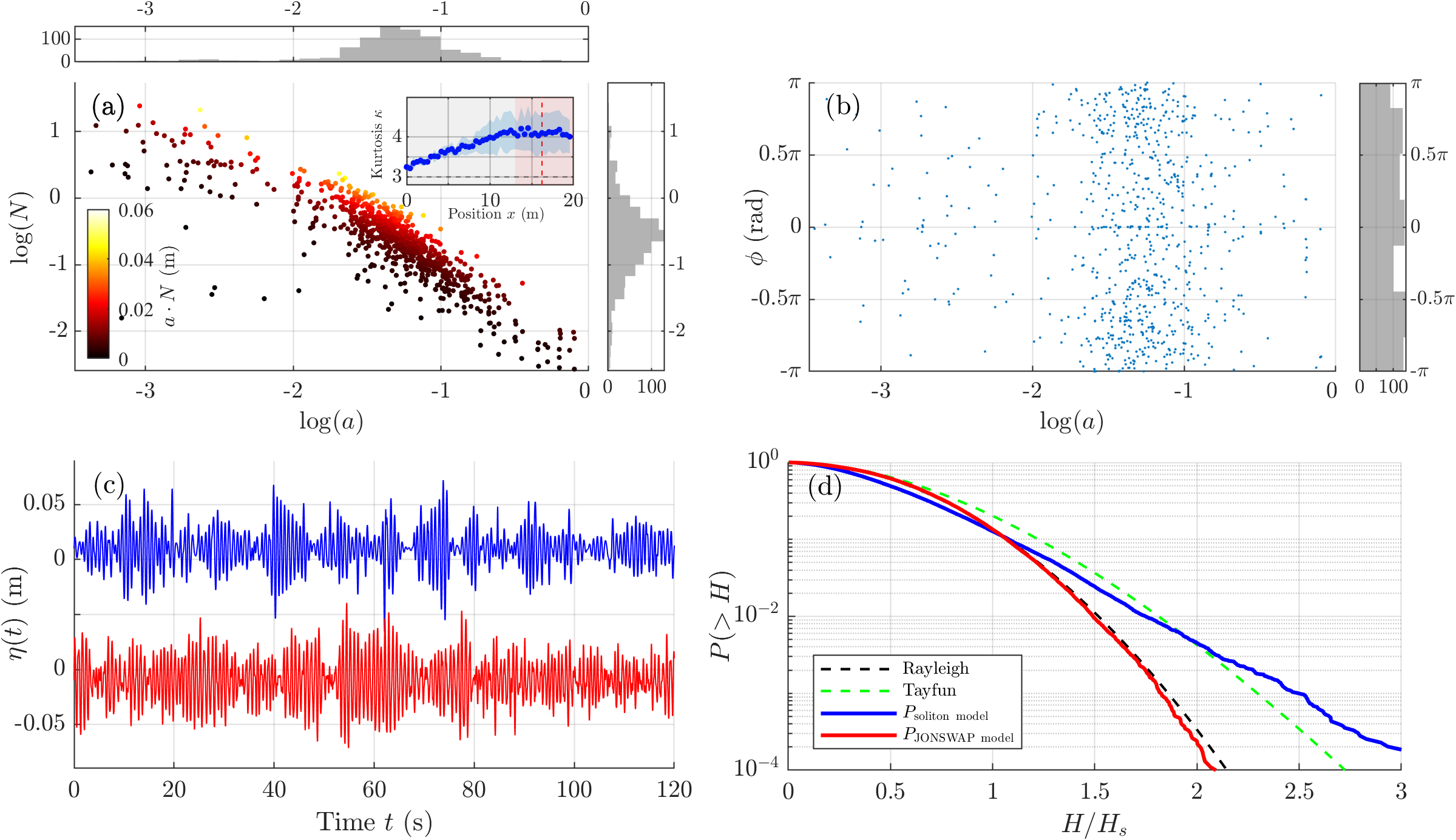}
\caption{Statistical properties of soliton parameters and JONSWAP ($\gamma=6$, $\varepsilon=0.08$, $f_p=1.3$ Hz) wave field reconstruction.
(a) Scatter plot of $\log(a)$ vs $\log(N)$ for 862 extracted solitons from a 1000-peak-period wave record at $x\approx 16$ m. Top and right panels show the marginal histogram distributions of $\log(a)$ and $\log(N)$, respectively, with color indicating the effective soliton amplitude $A=a\cdot N$. 
Top-right inset: Evolution of the kurtosis $\kappa$ along the wave flume. The blue dotted line denotes the mean value computed by segmenting the time-series at each gauge, and the shaded area indicates the $\pm 1\sigma$ uncertainty between 3 segments at each location. The red dashed line marks the reference measurement location $x\approx 16$ m, where the shaded red area determines the region of saturated kurtosis around a value of 4.
(b) Phase distribution $\phi$ of the extracted solitons, exhibiting a nearly uniform (white-noise) characteristic over $[-\pi, \pi]$. The right panel shows the marginal histogram of $\phi$.
(c) Qualitative comparison between the reconstructed wave field based on the log-normal $a-N$ joint distribution model with white-noise phase (blue) and the JONSWAP wave field (red) sharing identical significant wave height $H_s$, peak frequency $f_p$, and spectral bandwidth.
(d) Exceedance probability $P(>H)$ of 30,000-s synthetic time series generated by the JONSWAP model (red) and the novel soliton-based coherent structure model (blue), compared with the classical Rayleigh (black dashed) and Tayfun (green dashed) distributions.}
\label{fig2}
\end{figure*}

The statistical properties of the extracted solitons together with the subsequent reconstruction of random wave fields are shown in Figure~\ref{fig2}. In respective panel (a), the joint distribution of $\log(a)$ and $\log(N)$ exhibits a strong positive correlation, suggesting a power-law scaling $N\propto a^s$ modulated by multiplicative noise. Concurrently, the marginal Gaussian histograms signify that both parameters individually follow log-normal distributions. Such log-normal statistics seem to naturally emerge from the multiplicative nature of nonlinear wave interactions during the MI process. Crucially, these statistical features are entirely absent under linear wave conditions, establishing an analogy to the cascade processes observed in wave turbulence \cite{nazarenko2011wave}. This statistical observation provides strong experimental support for the log-normal law of soliton amplitude distributions, reinforcing the cascade analogy.

Figure~\ref{fig2}(b) displays the phase distribution $\phi$ of the extracted solitons, which closely follows a uniform distribution over $[-\pi,\pi]$. Indeed, these statistically uncorrelated phases distinguish the present coherent structure framework from deterministic breather models, underscoring the stochastic nature of the wave field.

Leveraging these statistical distributions, namely log-normal $a$ and $N$, uniform $\phi$, and uniform arrival times $t_0$, we generate a synthetic wave field via the superposition of these distinct solitons, a simplification justified by their temporal sparsity. Figure~\ref{fig2}(c) compares a representative segment of this reconstructed field (blue) against a standard JONSWAP wave field (red) with matching significant wave height $H_s$, peak frequency $f_p$, and spectral bandwidth. The two time series exhibit quantitative agreement in stochastic wave patterns, confirming that the soliton statistics effectively capture the essential dynamical and morphological features of a random sea state.

Figure~\ref{fig2}(d) compares the exceedance probabilities $P(>H)$ as a function of the wave height normalized by $H_s$. Notably, the soliton-based reconstruction (blue) exhibits a systematically heavier tail than the JONSWAP field (red). This disparity demonstrates that the superposition of coherent soliton structures inherently generates a higher probability of extreme waves than classical Gaussian random fields, even when bulk statistical parameters are identical. This heavy tail is a hallmark signature governed by the log-normal amplitude distribution coupled with the temporal sparsity of the solitons, a mechanism we quantify below.

Physically, the sparsity implies that each large-amplitude soliton acts as an approximately independent extreme event. The temporal separation suggests that one wave crest does not significantly influence the occurrence of the next. Consequently, the exceedance probability for the maximum wave crest height over a given observation period follows from extreme value theory, instead of the central limit theorem:
\begin{equation}\label{eqn-extreme value theory}
P_{\text{max}}(>H) = 1 - \exp\left[-\rho \cdot (1 - F_A(H))\right],
\end{equation}
where $\rho = \lambda T$ is the expected number of solitons (or independent wave events) during the observation time $T$, and $F_A(H)$ is the cumulative distribution function of individual soliton amplitudes $A = a \cdot N$.

In the limit of sparse soliton density for which we experimentally observe $\rho \approx 1$, i.e., the expected number of localized structures is around one per period, suggesting that the number of wave heights is comparable to that of the structures, the exponential can thus be approximated as $\exp[- (1-F_A)] \approx 1 - (1-F_A)$ for small $1-F_A$, yielding the simplified form 
\begin{equation}\label{eqn-extreme value theory simplified}
P(>H) \approx 1 - F_A(H), 
\end{equation}
with a note that this framework is suitable only for sparse structures. 

This limit corresponds to the physically relevant case in which the characteristic width of large-amplitude solitons is comparable to the mean wave period, ensuring that most solitons are well separated in time and contribute distinct wave crests. Under this condition, the exceedance probability $P(>H) = 1 - F_A(H)$ is interpreted as the probability that a randomly selected wave crest exceeds a given height $H$. For unidirectional waves examined in this work, $\rho \approx 1$ (arguably dependent on wave spectrum broadness) provides an excellent approximation, as confirmed by the experimental data.

\begin{figure}[htp]
\includegraphics[width=0.45\textwidth]{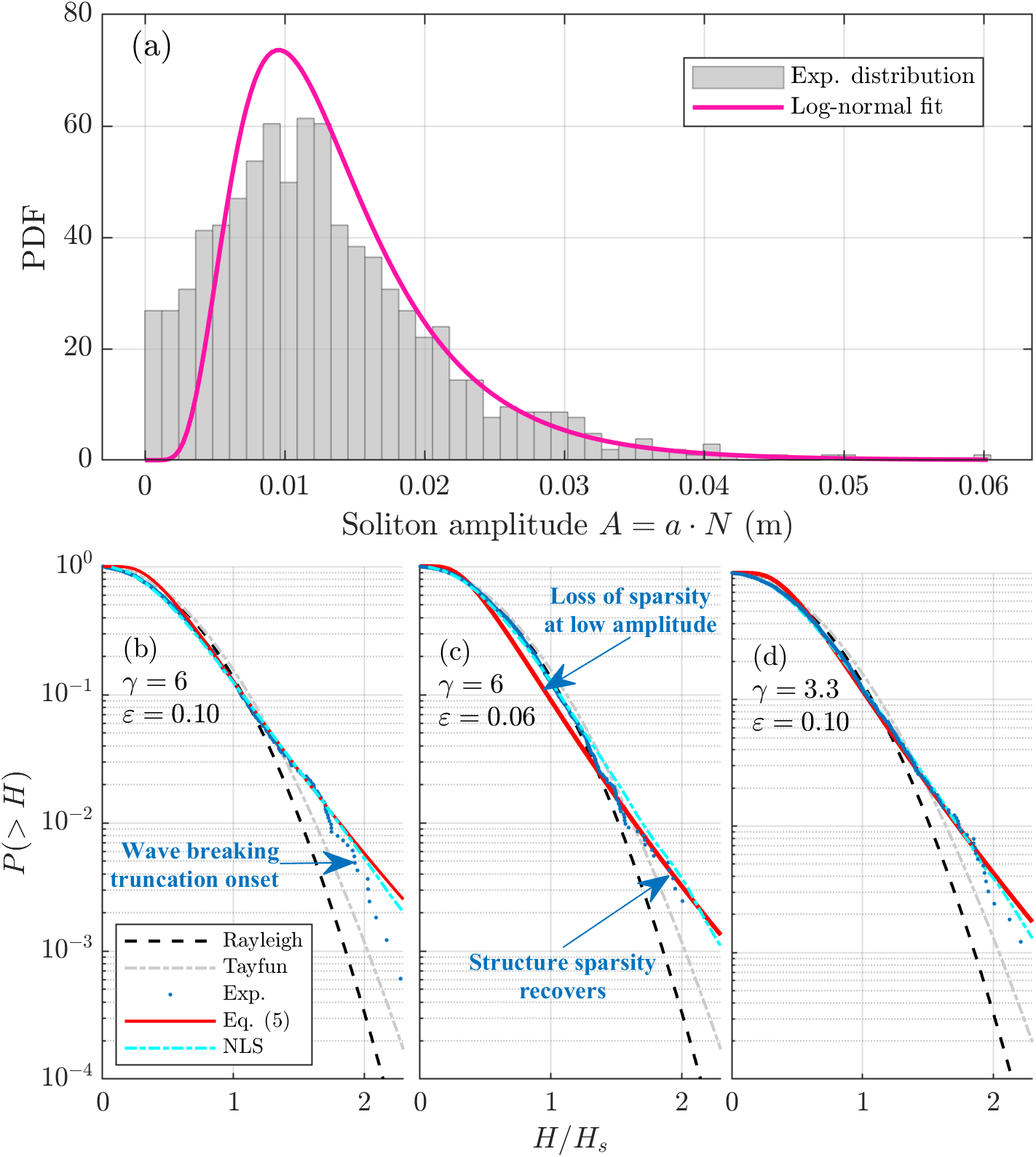}
\caption{Predictive theoretical framework for extreme wave probabilities. Experimental data recorded at $x\approx16$ m with saturated kurtosis value (blue dots) are compared against the simplified theory (Eq.~\ref{eqn-extreme value theory simplified}) mapped from the baseline statistics (red solid curve), the Tayfun distribution (gray dot-dashed), the Rayleigh distribution (black dashed), and the NLS simulation prediction for the respective sea states (cyan dot-dashed). 
(a) PDF with log-normal fit (\(\mu=-4.4\), \(\sigma=0.5\)) of effective soliton amplitudes \(A=a\cdot N\) extracted from the baseline experimental data as in Figure~\ref{fig1} with \(\varepsilon=0.08\) and peak-enhancement factor \(\gamma=6\). 
(b) Exceedance probability versus normalized wave height for an independent, higher-steepness JONSWAP sea state (\(\varepsilon=0.10\), \(\gamma=6\)).  
(c) Validation for a lower-steepness sea state (\(\varepsilon=0.06\), \(\gamma=6\)) based on (b) and comparing to the current theory (\(\mu=-4.5\), \(\sigma=0.5\)). 
(d) Validation for a steep and broader-band spectral sea state (\(\varepsilon=0.10\), \(\gamma=3.3\)) based on (b) and comparing to the proposed model (\(\mu=-4.4\), \(\sigma=0.48\)). 
}
\label{fig3}
\end{figure}

Note that $F_A(H)$ is the cumulative distribution function (CDF) of extreme soliton amplitudes $A=a\cdot N$, as shown in Fig.~\ref{fig3}(a), which experimentally turns out to be a log-normal distribution in the form of
\begin{equation}
F_A(H) = \Phi\!\left(\frac{\ln H - \mu}{\sigma}\right)\,
\end{equation}
where $\Phi$ denotes the standard normal CDF, while $\mu$ and $\sigma$ represent the mean and standard deviation of $\ln A$, respectively. Remarkably, this formula contains no adjustable parameters once $\mu$ and $\sigma$ are estimated from the data, yielding a direct and physics-based prediction of extreme events.

We now turn to the central result of this work by predicting extreme wave probabilities directly from the amplitude distribution of the coherent structures alone. The benchmarking data sets analyzed herein correspond to fully developed, strongly nonlinear wave fields characterized by a large kurtosis value of $\kappa\approx4$. This pronounced non-Gaussian regime indicates that nonlinear interactions have saturated, thereby providing a rigorous testbed for evaluating extreme wave predictability under saturated nonlinear dynamics.

Figure~\ref{fig3} validates this prediction against three independent experimental data sets. As demonstrated in panel (a), the extracted soliton amplitudes $A = a \cdot N$ from the baseline experimental data set adhere closely to a log-normal distribution. A slight discrepancy occurs in the low-amplitude region due to the high sensitivity of subdominant structures to background noise. However, this regime remains largely irrelevant to extreme wave statistics. Crucially, the distribution tail, which governs energy concentration and extreme wave emergence, is well-captured by the log-normal fit, thereby providing a reliable operational basis for the analytical extreme value theory developed below. Note that the mean and standard deviation of this log-normal shape may reasonably vary across different sea states. 

We evaluate this predictive capability against three distinct sea states in panels (b)-(d), respectively corresponding to the highly nonlinear sea state, the weakly nonlinear sea state, and a broadband sea state. To confirm the reliability of our observation and exclude possible randomness, the cyan dot-dashed lines in panels (b)-(d) represent predictions from NLS simulations of the respective sea states with approximately 10000 waves each, i.e., about ten times the length of the experimental data, and are indeed in excellent agreement with the data. 

Panel (b) first examines a more strongly nonlinear sea state with characteristic steepness $\varepsilon = 0.10$. The proposed theory from Eq.~\ref{eqn-extreme value theory simplified} exhibits excellent agreement with the experimental tail up to $H/H_s \leq  2$, corresponding to the $\rho = 1$ limit where solitons are sparse and well-separated. For the relevant extreme events ($H/H_s \geq 2$), the experimental probabilities fall below the theoretical prediction. This deviation is consistent with the physical constraint imposed by wave breaking and resulting energy loss, which we have characterized in our previous work~\cite{eeltink2022nonlinear}. Barring these breaking-induced effects, the overall agreement spans more than two orders of magnitude in probability.

Panel (c) considers a sea state with a lower steepness $\varepsilon=0.06$. Here, the reduced wave steepness leads to a slight decrease in the log-normal scale parameter, from $\mu = -4.4$ to $-4.5$, while the shape parameter $\sigma = 0.5$ remains unchanged. Remarkably, this single-parameter adjustment captures the extreme tail ($H/H_s \geq 1.5$) with high fidelity. The deviation observed for $H/H_s \leq 1.5$ is attributed to the broadening of soliton-like structures because of their lowered amplitudes, which deviates from the sparse distribution assumption for moderate-amplitude waves.

Panel (d) further tests the theory against a broader spectral bandwidth $\gamma=3.3$, while retaining the characteristic steepness $\varepsilon=0.10$, yielding the same $\mu = -4.4$ as in panel (b), but a slightly reduced shape parameter $\sigma = 0.48$. This mirrors the effect observed in panel (c). Specifically, the broader spectrum suppresses the efficiency of MI, resulting in less frequent coherent-structure interactions and thus reduced wave extremity, while the structural sparsity remains well preserved owing to the large structure amplitudes, as in panel (b).

 The fact that different physical mechanisms, that is, reduced steepness and broader bandwidth, manifest through adjustments of $\mu$ and $\sigma$ highlights the feasibility and physical interpretation of the sparse soliton framework. In fact, the value of $\mu$ encodes the overall amplitude scale of coherent structures, while the shape $\sigma$ reflects the effective degree of nonlinear organization. Both are determined directly from the structures' distribution, requiring almost no empirical fitting to the extreme tail. These results demonstrate that the sparse soliton framework provides a physically transparent prediction of extreme wave probabilities, comparing favourably with conventional models. The success of this prediction confirms that heavy tails originate from the intrinsic statistics of sparse coherent structures, complementing approaches based on higher-order perturbative corrections. 
Moreover, while statistical forecasts based on the NLS framework can qualitatively capture stochastic RW probabilities, it may significantly underestimate the amplitudes of individual highly steep RWs due to the limited order of approximation of the underlying water wave problem~\cite{slunyaev2013super}.

In summary, our results provide a direct mechanism-based link between a sparse coherent structure representation and RW statistics, offering a predictive framework that requires only the log-normal amplitude distribution of the extracted coherent structures as input. We emphasize that the envelope solitons employed here serve as templates for localized wave packets and do not necessarily propagate as persistent solitons within the random wave background; alternative localized templates may therefore be considered without altering the underlying concept of sparsity. Likewise, sparsity does not preclude nonlinear interactions between neighbouring structures, which may locally redistribute wave energy. The present framework should thus be regarded as a first attempt at a statistical description linking coherent structure amplitudes to extreme event probabilities, while a detailed treatment of their nonlinear interactions and an extension to directional seas remain subjects for future work. More broadly, this approach may provide a useful framework for analysing extreme event statistics in other nonlinear dispersive systems, including nonlinear optics and plasmas.

The authors would like to thank Prof. Jinghua Wang, Prof. Zeng Liu, and Dr. Lei Wang for helpful and constructive discussions. A.C. acknowledges support from Okinawa Institute of Science and Technology (OIST) with subsidy funding from the Cabinet Office, Government of Japan.

\bibliography{reference}
\clearpage
\newpage
\appendix
\onecolumngrid
\begin{center}
\section*{Supplemental Material}
\end{center}
\twocolumngrid

\section*{Iterative Matching Pursuit Algorithm for Soliton Decomposition}
In this appendix section, we provide a detailed description of the iterative matching pursuit algorithm used to decompose the measured wave envelope $\psi(t)$ into a sparse ensemble of soliton-like packets. This algorithm is designed to extract the coherent structures that govern the nonlinear dynamics, as discussed in the main text.

The soliton decomposition is achieved by representing the complex envelope as a superposition of localized soliton templates
\begin{equation}
\psi(t) = \sum_{j} N_j a_j \operatorname{sech}\left(\frac{\sqrt{2}}{2} a_j k_p \omega_p (t - t_{0,j})\right) e^{i\phi_j},
\end{equation}
where each template is defined by four parameters: the amplitude $a_j$, the continuous shape factor $N_j$, the carrier phase $\phi_j$, and the temporal center $t_{0,j}$. The iterative extraction process proceeds through the following steps:

\begin{itemize}

\item[] \textbf{Envelope Extraction:} The complex envelope $\psi(t)$ is extracted from the measured surface elevation $\eta(t)$ via bandpass filtering around the peak frequency $f_p$, followed by the Hilbert-Huang transform, i.e., $\psi(t)=\mathcal{H}\left[\eta\left(t\right)\right]$.
\item[]\textbf{Residual Initialization:} The residual signal is initialized as the full envelope, $r(t) = \psi(t)$.
\item[] \textbf{Best-Fit Soliton Search:} At each iteration, the algorithm identifies the most energetic localized wave packet within the residual. This is achieved by fitting the soliton template using an adaptive grid search over the parameters $a$, $N$, and $\phi$, followed by local refinement. The objective is to minimize a weighted error over a localized time window.
\item[]\textbf{Subtraction and Storage:} The dominant soliton is subtracted from the residual: $r(t) \leftarrow r(t) - \psi_j(t)$. The extracted parameters ${a_j, N_j, \phi_j, t_{0,j}}$ are stored in a dictionary.
\item[]\textbf{Convergence Criterion:} The procedure iterates until the energy of the residual signal drops below $5\%$ of the initial energy. The final output is the ensemble of all identified solitons.
\end{itemize}

\begin{figure}[htbp]
    \centering
    \begin{tikzpicture}[node distance=1.5cm, font=\small]
    
    \tikzstyle{startstop} = [rectangle, rounded corners, minimum width=2.2cm, minimum height=0.8cm, text centered, draw=black, fill=red!30]
    \tikzstyle{process} = [rectangle, minimum width=3.2cm, minimum height=0.8cm, text centered, draw=black, fill=blue!30]
    \tikzstyle{decision} = [diamond, minimum width=2.5cm, minimum height=0.8cm, aspect=2.5, text centered, draw=black, fill=green!30]
    \tikzstyle{arrow} = [thick, ->, >=stealth]
    
    \node (start) [startstop] {Start: Input \(\eta(t)\)};
    
    \node (hilbert) [process, below of=start] {Extract envelope \(\psi(t)\) via Hilbert-Huang};
    
    \node (init) [process, below of=hilbert] {Initialize residual \(r(t) = \psi(t)\)};
    
    \node (search) [process, below of=init, align=center] {Adaptive grid search for best-fit soliton: \\
    \(\{a, N, \phi, t_0\}\)};
    
    \node (subtract) [process, below of=search] {Subtract soliton: \(r(t) \leftarrow r(t) - \psi_j(t)\)};
    
    \node (energy) [decision, below of=subtract, align=center] {Residual energy \\ \(< 5\%\)?};
    
    \node (store) [process, left of=energy, node distance=4.5cm, align=center] {Store soliton params: \\
    \(\{a_j, N_j, \phi_j, t_{0,j}\}\)};
    
    \node (end) [startstop, below of=energy, node distance=1.8cm] {End: Output ensemble};
    
    \draw [arrow] (start) -- (hilbert);
    \draw [arrow] (hilbert) -- (init);
    \draw [arrow] (init) -- (search);
    \draw [arrow] (search) -- (subtract);
    \draw [arrow] (subtract) -- (energy);
    
    \draw [arrow] (energy.south) -- node[right, font=\footnotesize] {Yes} (end.north);
    
    \draw [arrow] (energy.west) -- node[above, font=\footnotesize] {No} (store.east);
    \draw [arrow] (store.west) -- ++(-0.5,0) |- ([yshift=0.6cm]search.north) -- (search.north);
    
    \end{tikzpicture}
    \caption{Flowchart of the iterative matching pursuit algorithm for extracting sparse soliton-like packets from the measured wave envelope. The loop continues until the residual energy falls below the 5\% threshold.}
    \label{fig:algorithm_flowchart}
\end{figure}
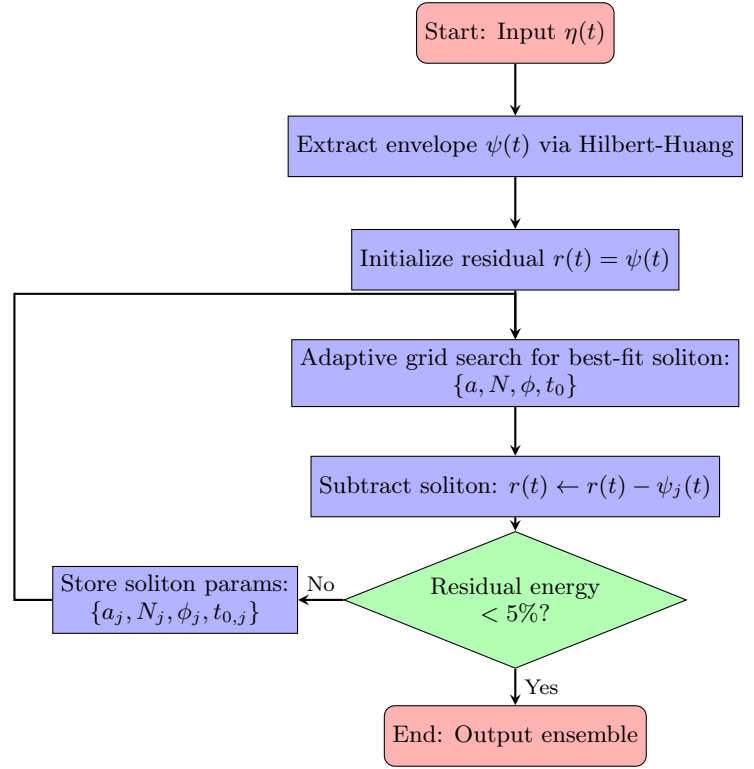

A schematic representation of this algorithm is provided in Fig.~\ref{fig:algorithm_flowchart}.

For completeness, we outline the pseudo-code implementation of the soliton decomposition procedure

\begin{itemize}
    \item[] Extract envelope \(\psi(t)\) from \(\eta(t)\).
    \item[] Set iteration index \(j = 1\) and residual \(r^{(1)}(t) = \psi(t)\).
    \item[] While \(\|r^{(j)}(t)\|^2 / \|\psi(t)\|^2 > 0.05\):
    \begin{itemize}
        \item[] Find the best-fit soliton parameters \(\{a_j, N_j, \phi_j, t_{0,j}\}\) that minimize the weighted local error between \(r^{(j)}(t)\) and the soliton template.
        \item[] Construct the soliton template \(\psi_j(t)\) using Eq.~(1).
        \item[] Update the residual: \(r^{(j+1)}(t) = r^{(j)}(t) - \psi_j(t)\).
        \item[] Store the parameters \(\{a_j, N_j, \phi_j, t_{0,j}\}\).
        \item[] Increment \(j \leftarrow j + 1\).
    \end{itemize}
    \item[] Output the complete dictionary of soliton parameters.
\end{itemize}
This algorithm ensures that the extracted basis functions are adapted to the nonlinear coherent structures present in the wave field, rather than imposed by an arbitrary mathematical criterion. As mentioned in the manuscript, the physical fidelity of this representation is validated by the excellent agreement between the reconstructed envelope and the experimental data, as demonstrated in Fig.~1(b).

\section*{Random peak emergence times of coherent structures}
\begin{figure}[htp]
\centering
\includegraphics[width=0.8\columnwidth]{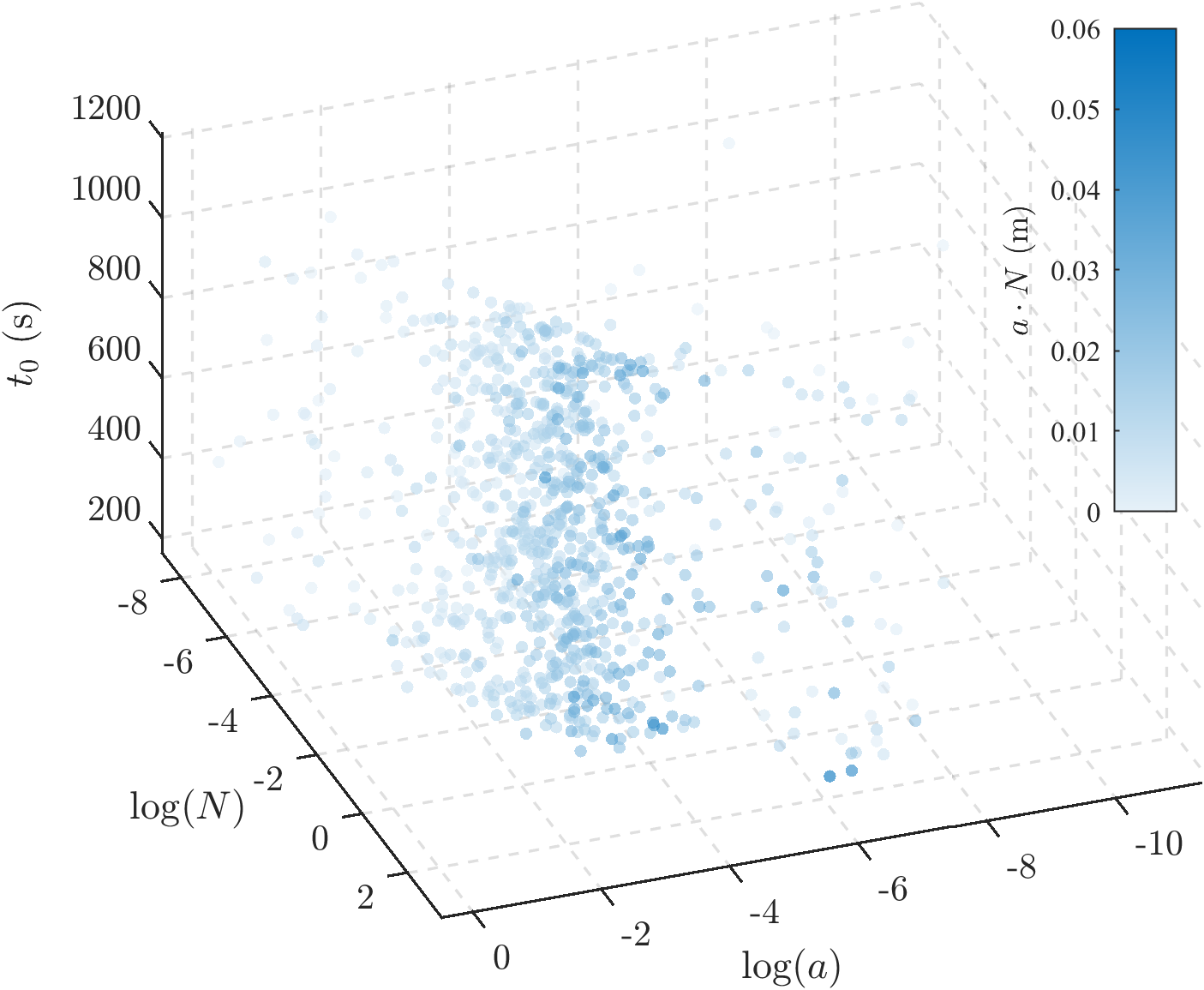}
\caption{Three-dimensional scatter plot of extracted soliton parameters: $\log(a)$ versus $\log(N)$ versus peak emergence time $t_0$. The points are distributed uniformly along the $t_0$ axis over the full measurement duration, indicating that the coherent soliton-like packets emerge randomly in time with no preferred temporal localization.}
\label{fig:appendix_arrival_time}
\end{figure}

Following the soliton decomposition described above, we examine the statistical properties of the extracted soliton parameters. Of particular interest is the distribution of peak emergence times $t_0$ of the individual soliton components. Note that this is not shown in Figure~\ref{fig2}(a). Figure~\ref{fig:appendix_arrival_time} further depicts a three-dimensional scatter plot of the extracted soliton parameters: $\log(a)$ versus $\log(N)$ versus $t_0$, where $a$ is the soliton amplitude scaling factor, $N$ is the effective soliton order, and $t_0$ is the peak emergence time.

The scatter points are distributed uniformly along the $t_0$ axis over the entire measurement window. This observation is consistent with the interpretation that the coherent soliton-like packets emerge as independent events without preferred temporal localization, aside from the natural clustering associated with wave group formation. The uniform distribution of $t_0$ further supports the sparse coherent structure framework adopted in this work, where the random wave field is described as a superposition of localized packets with random phases and peak emergence times.
\end{document}